\documentclass{emulateapj} 
\usepackage{graphicx}
\usepackage{amssymb,amsfonts,amsmath,amstext,amsgen,amsopn,amsxtra,indentfirst,lscape,times,rotating}

\begin{document}

\title{The Influence of Atmospheric Scattering and Absorption \\ on Ohmic Dissipation in Hot Jupiters}
\author{Kevin Heng\altaffilmark{1,2}}
\altaffiltext{1}{ETH Z\"{u}rich, Institute for Astronomy, Wolfgang-Pauli-Strasse 27, CH-8093, Z\"{u}rich, 
Switzerland}
\altaffiltext{2}{Zwicky Prize Fellow}

\begin{abstract}
Using semi-analytical, one-dimensional models, we elucidate the influence of scattering and absorption on the degree of Ohmic dissipation in hot Jovian atmospheres.  With the assumption of Saha equilibrium, the variation in temperature is the main driver of the variations in the electrical conductivity, induced current and Ohmic power dissipated.  Atmospheres possessing temperature inversions tend to dissipate most of the Ohmic power superficially, at high altitudes, whereas those without temperature inversions are capable of greater dissipation deeper down.  Scattering in the optical range of wavelengths tends to cool the lower atmosphere, thus reducing the degree of dissipation at depth.  Purely absorbing cloud decks (in the infrared), of a finite extent in height, allow for localized reductions in dissipation and may reverse a temperature inversion if they are dense and thick enough, thus greatly enhancing the dissipation at depth.  If Ohmic dissipation is the mechanism for inflating hot Jupiters, then variations in the atmospheric opacity (which may be interpreted as arising from variations in metallicity and cloud/haze properties) and magnetic field strength naturally produce a scatter in the measured radii at a given strength of irradiation.  Future work will determine if these effects are dominant over evolutionary effects, which also contribute a scatter to the measured radii.
\end{abstract}


\section{Introduction}

A widely-accepted belief circulating within the astrophysical literature is that the intense starlight impinging upon hot Jovian atmospheres liberates electrons from their Group {\sc i} metals (potassium and sodium), thus providing a source of free electrons within the atmospheric flow \citep{bs10,pmr10b}.  The balance of collisional ionization with recombination allows for the application of the Saha equation towards estimating the ionization fraction of the flow.  If an ambient magnetic field exists, its intersection by the $\sim 1$ km s$^{-1}$ flow produces induced currents which act to oppose the flow itself --- a global, exoplanetary-scale manifestation of Lenz's law.  The mostly neutral flow is coupled to the electrons via collisions, and the opposing forces act to retard the flow, a process known as magnetic drag.  Ohmic dissipation converts the mechanical energy of the flow into heat, which acts to delay or even suspend the contraction of the exoplanet \citep{wu12}.  Thus, hot Jovian atmospheres behave like giant electrical circuits.

It has been observed that hot Jupiters have radii which increase with the strength of irradiation, with a non-negligible scatter present in the measured radii at a given strength of irradiation \citep{burrows07,enoch11,laughlin11,mf11,ds11}.  (See also Figure 8 of \citealt{php12}.)  A previously overlooked aspect in theoretical considerations of Ohmic dissipation is that the variations of atmospheric scattering and absorption account naturally for this observed scatter, including the absence or presence of a temperature inversion in the atmosphere \citep{burrows07b,fortney08}.  These variations in opacity may be attributed to variations in metallicity and/or the properties of clouds/hazes (e.g., \citealt{burrows11,helling11}), the presence of which has been inferred from transit observations obtained using the \textit{Hubble Space Telescope} \citep{pont08,sing11,gibson12}.

In this Letter, we demonstrate these statements by using simple, semi-analytical models of temperature-pressure profiles \citep{guillot10,hhps12} to compute the Ohmic dissipation for a variety of model atmospheres.  With the benefit of hindsight gleaned from three-dimensional simulations of atmospheric circulation \citep{php12}, we adopt simplifying assumptions for the atmospheric dynamics and focus on the variations in temperature, since the exponential dependence of the ionization fraction on temperature is the overwhelming factor in determining the degree of Ohmic dissipation as well as how deeply it penetrates into the atmosphere.  We describe our results in \S\ref{sect:method}, discuss their implications in \S\ref{sect:discussion} and summarize our conclusions in \S\ref{sect:conclusion}.

\begin{figure}
\centering
\includegraphics[width=\columnwidth]{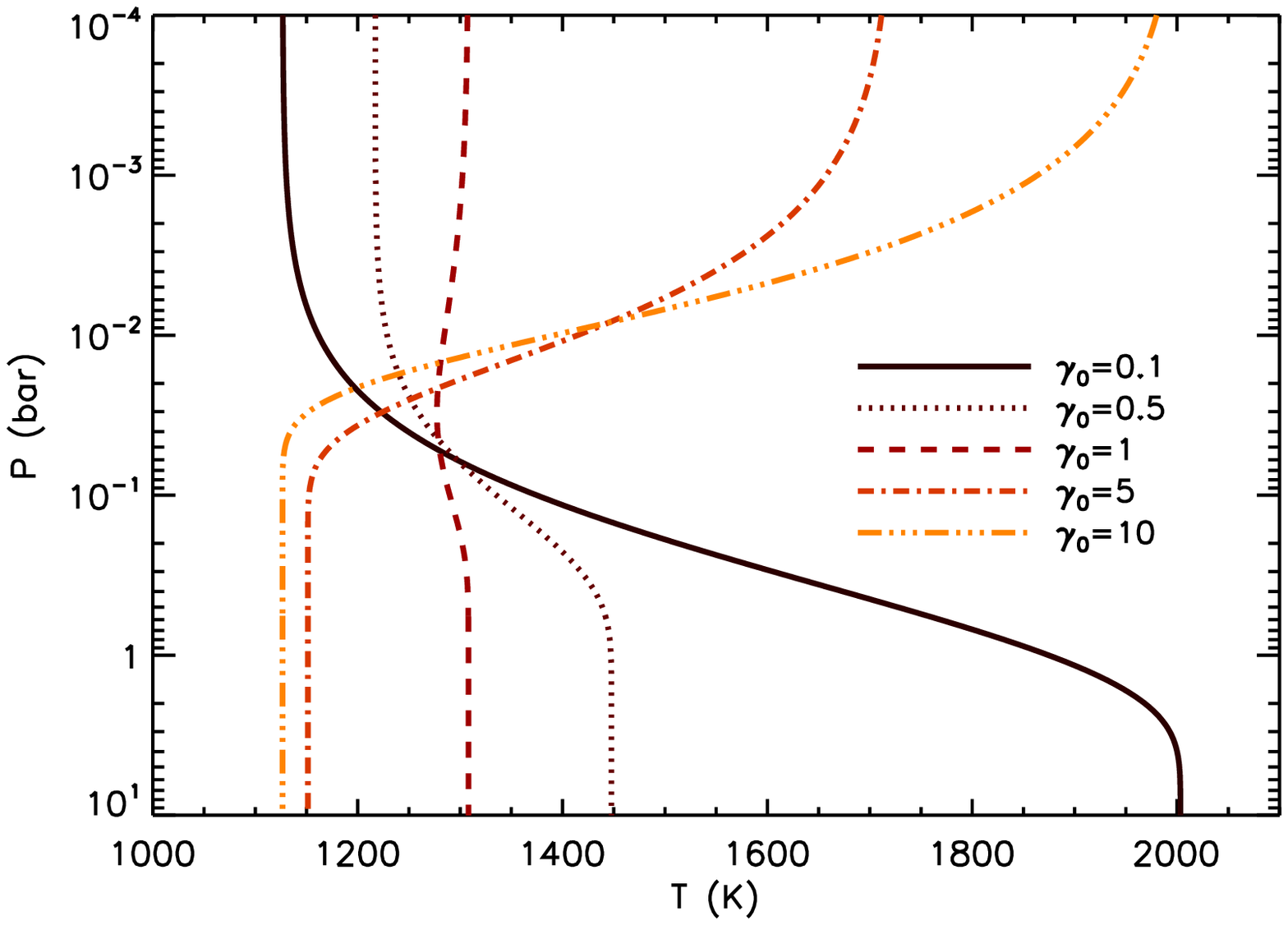}
\includegraphics[width=\columnwidth]{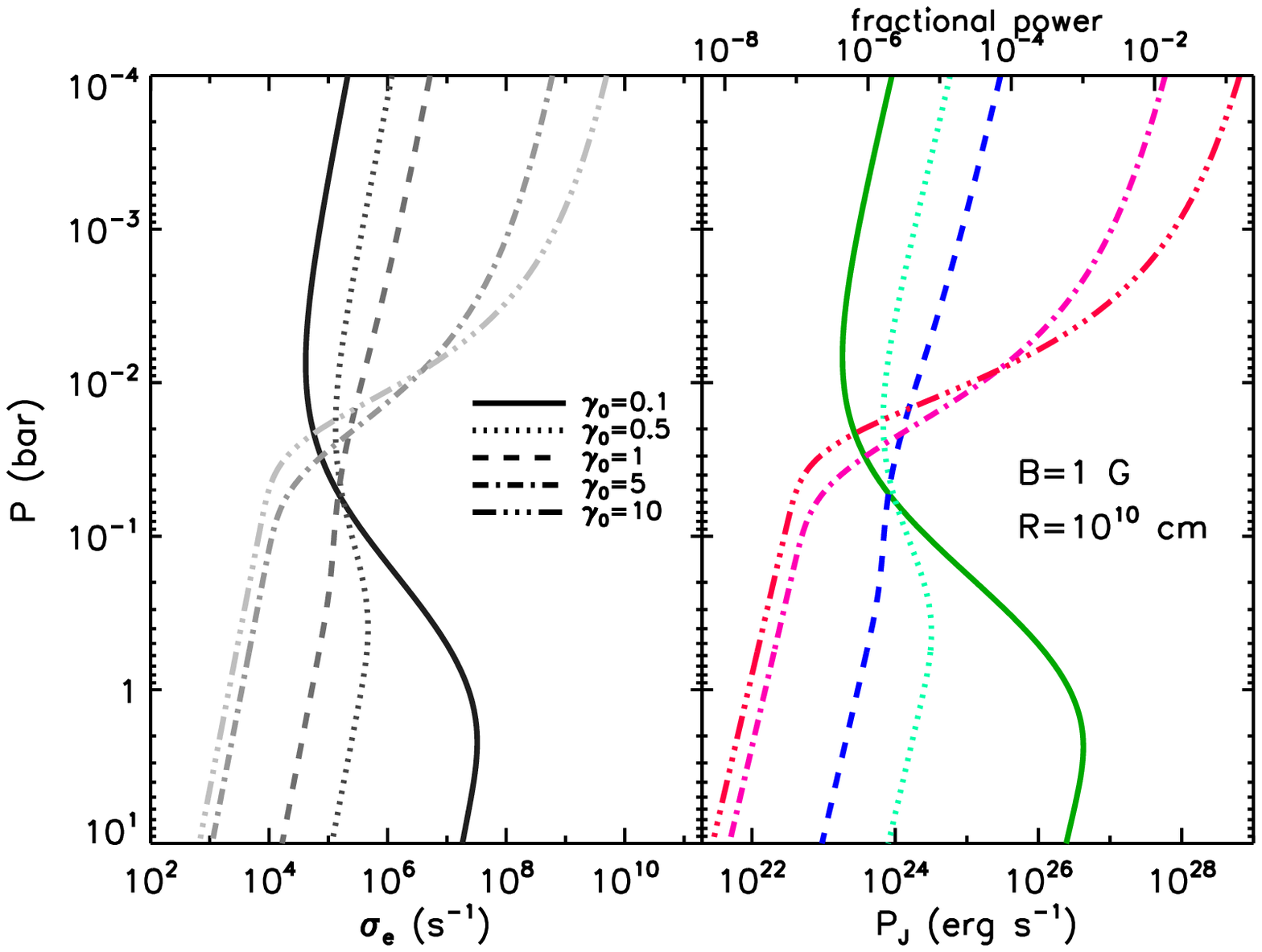}
\caption{Top panel: temperature-pressure profiles used in this study, where the ratio of shortwave to longwave opacities ($\gamma_0$) is varied.  The irradiation temperature of all of these profiles has been set to $T_{\rm irr} = 1850$ K.  Bottom panel: variation with pressure/height of the electrical conductivity (bottom left panel) and Ohmic power dissipated (bottom right panel) for models with $\gamma_0=0.1$--10.  For the bottom right panel, the horizontal axis at the top is the Ohmic power normalized by the incident stellar luminosity.}
\label{fig:tp}
\end{figure}

\section{Methodology}
\label{sect:method}

We assume a purely gaseous atmosphere dominated by molecular hydrogen and possessing an ambient magnetic field of unspecified geometry with a strength $B$.

\subsection{Time Scales}

Magnetic drag and Ohmic dissipation are effective processes only if the electrons and hydrogen molecules are tightly coupled, such that when the ambient magnetic field acts on the former, the effect is also felt by the latter.  For typical parameter values, the collisional time scale \citep{draine83},
\begin{equation}
t_{\rm coll} \approx 10^{-10} \mbox{ s} ~\left( \frac{\rho}{10^{-6} \mbox{ g cm}^{-3}} \right)^{-1} \left( \frac{T}{1500 \mbox{ K}} \right)^{-1/2},
\label{eq:tcoll}
\end{equation}
is the shortest one in the system, where $\rho$ denotes the mass density and $T$ the temperature.  In order for the magnetic effects to be communicated throughout the flow, we require $t_{\rm coll} < t_{\rm gyro}$, where $t_{\rm gyro} = 2\pi m_e c/e B \sim 10^{-7}$ s $(B/1\mbox{ G})^{-1}$ is the time scale associated with the gyration of an electron, with mass $m_e$ and charge $e$, about a magnetic field line.  (The speed of light is denoted by $c$.)  This condition is trivially satisfied,
\begin{equation}
B < 3.4 \mbox{ kG} ~\left( \frac{\rho}{10^{-6} \mbox{ g cm}^{-3}} \right) \left( \frac{T}{1500 \mbox{ K}} \right)^{1/2}.
\end{equation}
If the gyration time scale is shorter than the collisional time scale, then the opposing forces acting on the electrons are not communicated to the neutrals, but this requires implausible values for the magnetic field strength.

Magnetic field lines anchored to the deep interior of a hot Jupiter may be perturbed by the atmospheric flow, across a pressure scale height $H=kT/mg$, with the perturbations being communicated at the Alfv\'{e}n speed $v_{\rm A} = B/2(\pi \rho)^{1/2}$.  The assumption of a steady-state, ambient magnetic field is reasonable if the Alfv\'{e}n time scale ($t_{\rm A} \sim H/v_{\rm A}$) is less than the advective time scale associated with the zonal (east-west) atmospheric winds ($t_{\rm adv} \sim R/c_s$ where $R$ is the radius of the hot Jupiter and $c_s$ is the sound speed), which results in a lower bound for the magnetic field strength,
\begin{equation}
\begin{split}
B >& 0.20 \mbox{ G} ~\left( \frac{\rho}{10^{-6} \mbox{ g cm}^{-3}} \right)^{1/2} \left( \frac{T}{1500 \mbox{ K}} \right)^{3/2} \\
&\times \left( \frac{g}{10 \mbox{ m s}^{-2}} \right)^{-3/2} \left( \frac{R}{10^{10} \mbox{ cm}} \right)^{-1}.
\end{split}
\end{equation}

We note that in all of the preceding estimates, the mass density is $\rho \sim mg/kT \kappa_0 \sim 10^{-6}$ g cm$^{-3}$ $(T/1500 \mbox{ K})^{-1} (\kappa_0/0.01 \mbox{ cm}^2 \mbox{ g}^{-1})^{-1}$, where $m = 2 m_{\rm H}$ is the mean molecular mass, $m_{\rm H}$ is the mass of the hydrogen atom, $g=10$ m s$^{-2}$ is the surface gravity and $k$ is the Boltzmann constant.  The quantity $\kappa_0$ is the broadband absorption opacity in either the optical or the infrared range of wavelengths.

\subsection{Model Description}

\begin{figure}
\centering
\includegraphics[width=\columnwidth]{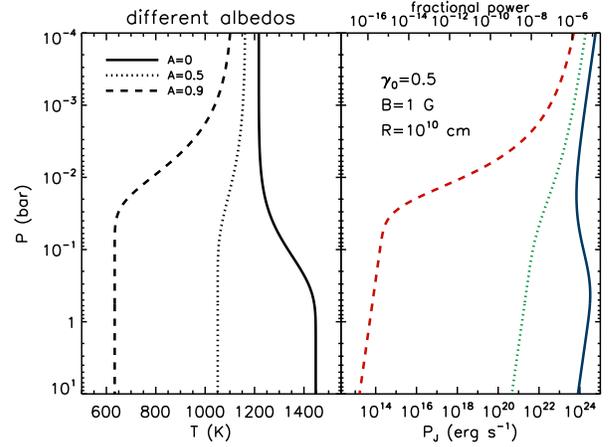}
\caption{Models with the same value of $\gamma_0=0.5$ but with different values of the Bond albedo ${\cal A}$.  Left panel: temperature-pressure profiles.  Right panel: Ohmic power dissipated for $B=1$ G.  For the right panel, the horizontal axis at the top is the Ohmic power normalized by the incident stellar luminosity.}
\label{fig:tp2}
\end{figure}

The incident flux impinging upon the substellar point is ${\cal F}_0 = \sigma_{\rm SB} T^4_{\rm irr}$, where $\sigma_{\rm SB}$ is the Stefan-Boltzmann constant and the irradiation temperature is
\begin{equation}
T_{\rm irr} = T_\star \left( \frac{R_\star}{a} \right)^{1/2} \left( 1 - {\cal A} \right)^{1/4}.
\end{equation}
Here, the stellar radius and effective temperature are represented by $R_\star$ and $T_\star$, respectively, while $a$ is the spatial separation between the exoplanet and the star.  We adopt $T_{\rm irr} = 1850$ K, which is a value midway between that of HD 189733b and HD 209458b, corresponding to ${\cal F}_0 \approx 6.6 \times 10^8$ erg cm$^{-2}$ s$^{-1}$ $(1-{\cal A})^{1/4}$; on Earth, ${\cal F}_0 \approx 1370$ W m$^{-2}$ is the solar constant (if the Bond albedo ${\cal A}$ is set to zero).  The incident stellar luminosity is ${\cal L}_0 = 2 \pi R^2 {\cal F}_0$; for $R = 10^{10}$ cm, we get ${\cal L}_0 \approx 4.2 \times 10^{29}$ erg s$^{-1}$.

For the radiative transfer, we adopt the dual-band approximation, which assumes that incident starlight and thermal emission from the hot Jupiter occur at distinct wavelengths (e.g., \citealt{guillot10,hfp11,hhps12}).  Atmospheric absorption is described by two broadband opacities: $\kappa_{\rm S}$ in the optical and $\kappa_{\rm L}$ in the infrared.  Their ratio is denoted by $\gamma_0 = \kappa_{\rm S}/\kappa_{\rm L}$.  Atmospheric scattering in the optical is described by the Bond albedo ${\cal A}$, albeit in the two-stream, collimated beam approximation.  We ignore the effect of collision-induced absorption, although we note that such an effect increases the temperature, and hence the Ohmic dissipation, at depth and may be included via the models of \cite{hhps12} as an additional parameter.  We set $\kappa_{\rm L} = 0.01$ cm$^2$ g$^{-1}$ such that the infrared photosphere resides at $\sim 0.1$ bar.  The broadband, optical opacity then follows from the assumed value of $\gamma_0$, i.e., $\kappa_{\rm S} = \gamma_0 \kappa_{\rm L}$.

Dynamically, the zonal (east-west) wind speed $v$ is the dominant component, such that the induced current density is primarily in the meridional (north-south) direction,
\begin{equation}
J \sim \frac{B v \sigma_e}{c},
\end{equation}
where $\sigma_e$ is the electrical conductivity.  We adopt the expression used in \cite{pmr10b} for $\sigma_e$ and set the potassium abundance to be solar ($a_{\rm K}=10^{-7}$).  The Ohmic power dissipated in each layer of the atmosphere, with a thickness of one pressure scale height $H$, is
\begin{equation}
P_{\rm J} \approx \frac{4 \pi R^2 H J^2}{\sigma_e} \sim \frac{4 \pi \Gamma}{g} \left( \frac{B R kT}{m c} \right)^2 \sigma_e,
\end{equation}
with $\Gamma=7/5$ denoting the adiabatic gas index for an ideal diatomic gas.  We have assumed that the zonal velocity is equal to the sound speed, i.e., $v=c_s$ where $c_s = ( \Gamma kT/ mg)^{1/2}$.  We truncate our models at $P=10$ bar, since simulations of atmospheric circulation have shown that strong zonal winds, with speeds of $v \sim 1$ km s$^{-1}$, penetrate down to this depth for the strength of stellar irradiation we have chosen (e.g., \citealt{hmp11,hfp11,php12}).  Since the zonal wind speeds approach or exceed the sound speed in these simulations, our assumption of $v=c_s$ is a reasonable one.

\subsection{Results}

The top panel of Figure \ref{fig:tp} shows the temperature-pressure profiles for model atmospheres with various values of the optical to infrared opacity ratio, $\gamma_0 = 0.1$--10.  Atmospheres with $\gamma_0 > 1$ possess temperature inversions, while those with $\gamma_0 < 1$ do not \citep{hubeny03,hansen08,guillot10}.  Even with this broad range in $\gamma_0$ values, the variations in temperature are modest, differing by less than a factor of 2.  By contrast, the electrical conductivity spans several orders of magnitude, as shown in the bottom left panel of Figure \ref{fig:tp}, due to its exponential dependence on temperature via the Boltzmann factor.  Atmospheres with temperature inversions tend to concentrate Ohmic power at high altitudes ($\lesssim 0.01$ bar), while those without temperature inversions tend to dissipate more power deeper down (bottom right panel of Figure \ref{fig:tp}).  For our model with $\gamma_0=0.1$, the local Ohmic dissipation at $P=10$ bar approaches 0.1\% of the incident stellar luminosity for $B=1$ G; this value scales with $B^2$.  For the most highly irradiated hot Jupiters, the $B^2$ scaling is expected to saturate due to the effects of magnetic drag, which act to halt the zonal winds and thus reduce the Ohmic dissipation \citep{menou12}.

We next examine the effects of scattering in the optical range of wavelengths.  The left panel of Figure \ref{fig:tp2} shows the temperature-pressure profiles for models with $\gamma_0=0.5$ and three different values of the Bond albedo: ${\cal A}=0, 0.5$ and 0.9.  The right panel of Figure \ref{fig:tp2} shows the corresponding Ohmic power dissipated, as a function of pressure, for $B=1$ G.  Shortwave scattering produces two effects.  The first and most obvious effect is to reduce the strength of the stellar irradiation via the diminution of $T_{\rm irr}$.  The second effect is to cool the lower atmosphere and thereby reduce the Ohmic dissipation at depth.

\begin{figure}
\centering
\includegraphics[width=\columnwidth]{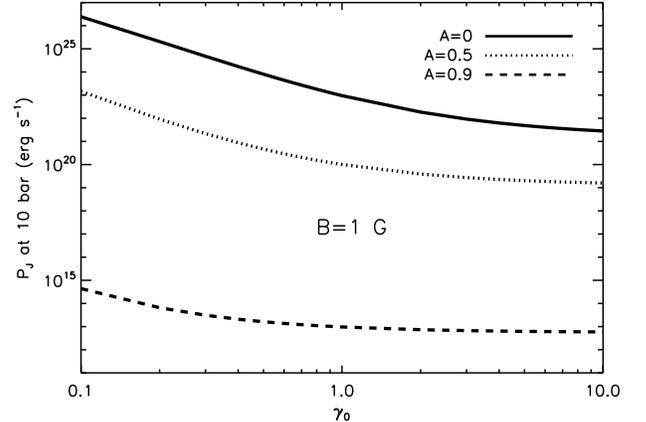}
\caption{Ohmic power dissipated at $P=10$ bar for models with $\gamma_0=0.1$--10 and ${\cal A}=0, 0.5$ and 0.9.  The assumed magnetic field strength is $B=1$ G.}
\label{fig:depth}
\end{figure}

\cite{gs02} have suggested that power dissipated at $\sim 10$ bar may affect the evolution of a hot Jupiter and provide an explanation for its inflated radius.  (See also \citealt{burrows03} and \citealt{wu12}.)  In Figure \ref{fig:depth}, we show the Ohmic power dissipated at $P=10$ bar for our models with $\gamma_0=0.1$--10 and ${\cal A}=0, 0.5$ and 0.9, again assuming $B=1$ G.  In addition to introducing a spread in the Ohmic power dissipated at a given $\gamma_0$ value, shortwave scattering also reduces the difference in $P_{\rm J}$ between low- and high-$\gamma_0$ models for a given value of ${\cal A}$.

\begin{figure}
\centering
\includegraphics[width=\columnwidth]{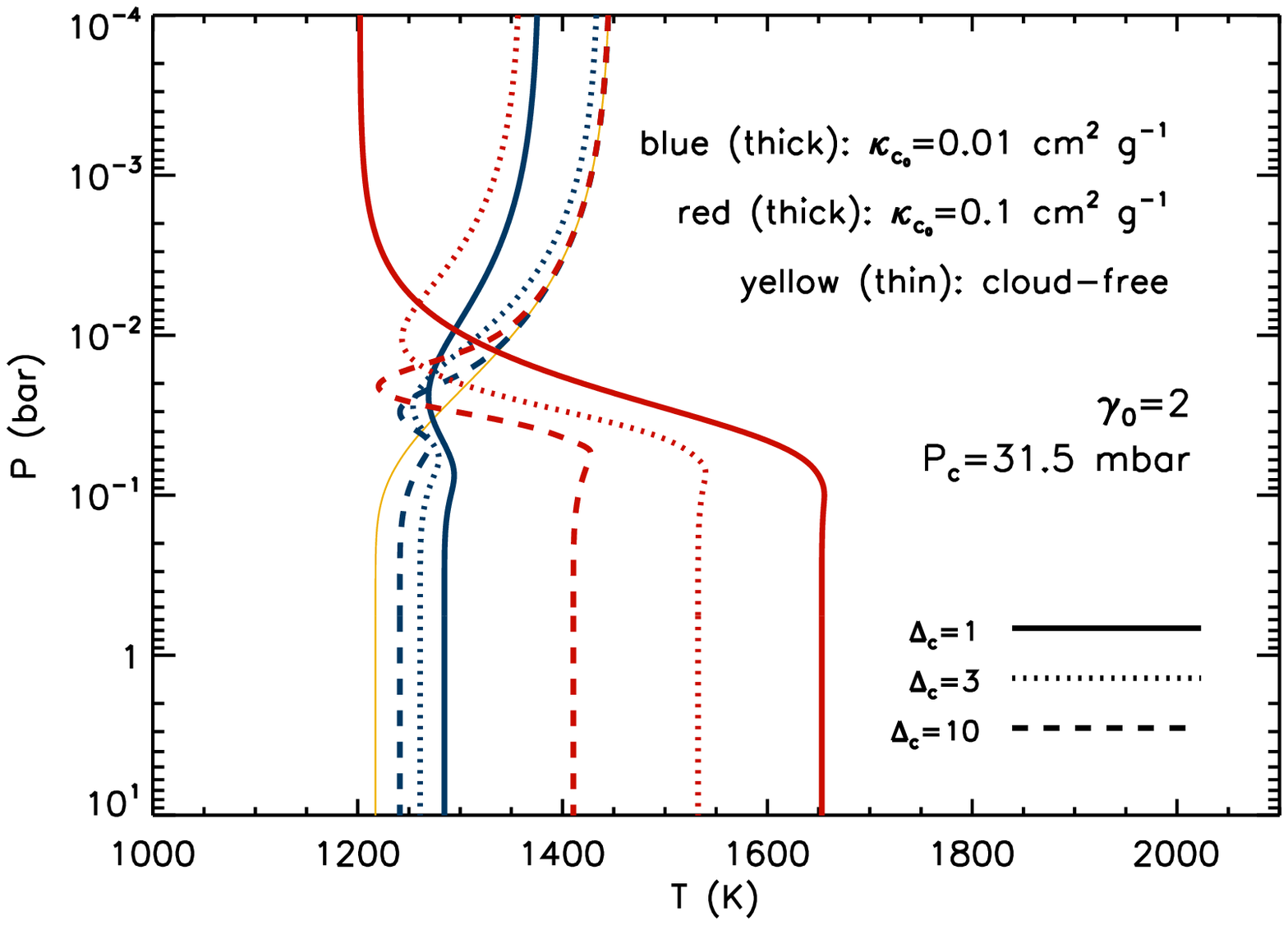}
\includegraphics[width=\columnwidth]{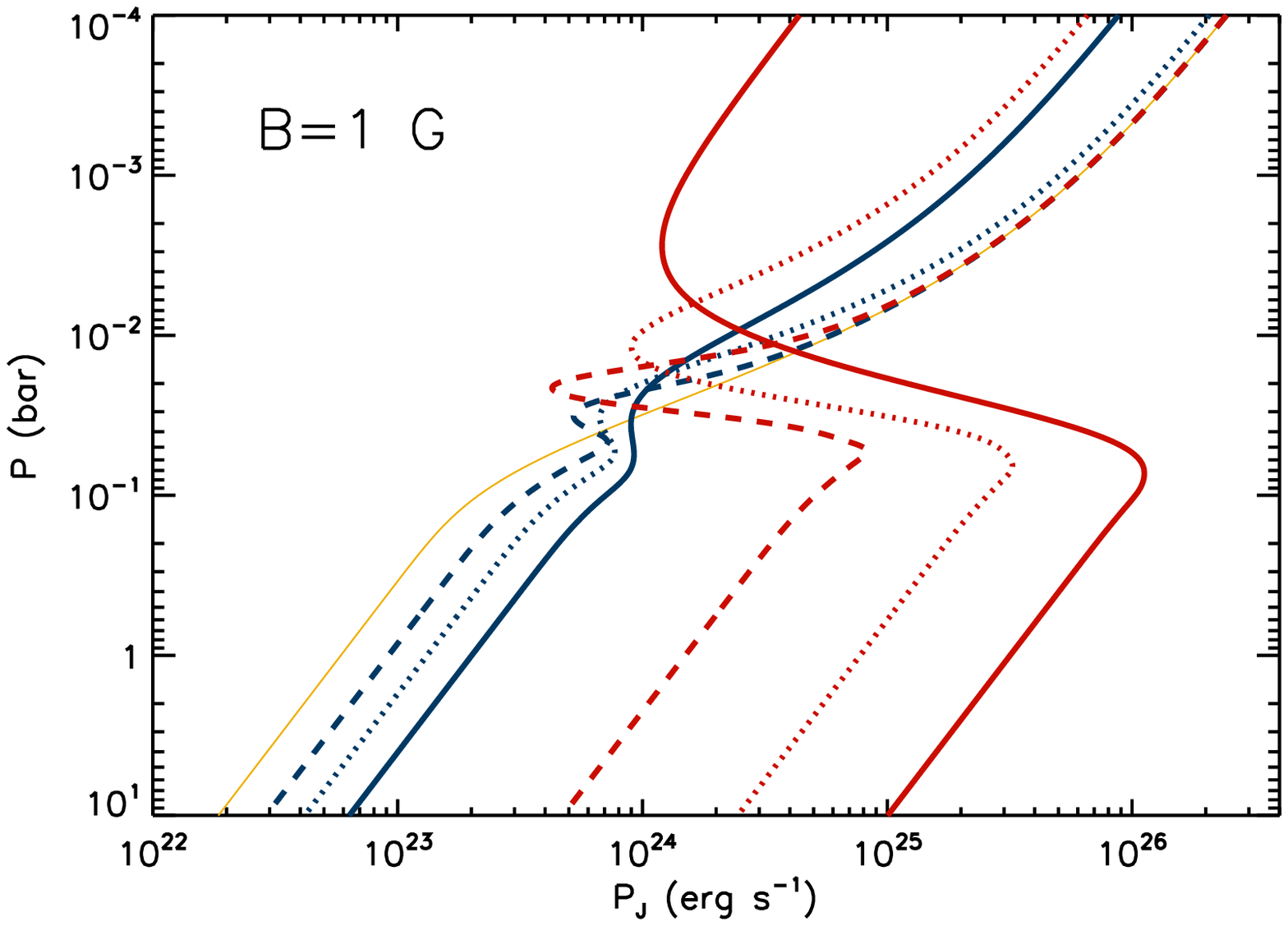}
\caption{Models with a Gaussian cloud deck, absorbing only in the infrared, centered at $P_{\rm c}=0.01$ bar.  Top panel: temperature-pressure profiles which include cloud decks with different absorption opacities and thicknesses.  Bottom panel: Ohmic power dissipated for $B=1$ G.  In both panels, the thin, yellow curve shows the cloud-free case for $\gamma_0=2$.}
\label{fig:cloud}
\end{figure}

Finally, we consider purely-absorbing cloud decks (in the infrared) with a Gaussian shape/profile (Figure \ref{fig:cloud}), as described in \cite{hhps12}.  We examine dense ($\kappa_{\rm c_0}=0.1$ cm$^2$ g$^{-1}$) versus tenuous ($\kappa_{\rm c_0}=0.01$ cm$^2$ g$^{-1}$) and thick ($\Delta_{\rm c}=1$) versus thin ($\Delta_{\rm c}=10$) cloud decks.  In all of the models shown, the photon deposition depth (in the absence of scattering) is located at $P_{\rm D} \approx 0.63 g/\kappa_{\rm S} = 31.5$ mbar.  We have intentionally centered the cloud decks at $P_{\rm c}=31.5$ mbar, as it has been shown by \cite{hhps12} that this maximizes the effects we are about to discuss.  Cloud decks absorbing purely in the infrared produce a greenhouse effect, warming the lower atmosphere and cooling the upper atmosphere, with denser clouds producing a greater effect.  If the cloud deck is thin enough, it imprints its shape onto the temperature-pressure profile, thus producing localized reductions in the Ohmic power dissipated.  If the cloud deck is dense and thick enough ($\Delta_{\rm c}=1$), then it reverses the temperature inversion altogether, thus greatly enhancing the Ohmic dissipation at depth.  If the cloud deck sits above the photon deposition depth ($P_{\rm c} < P_{\rm D}$), then these effects are diminished.  If it sits below the photon deposition depth ($P_{\rm c} > P_{\rm D}$), then it cools the upper atmosphere but has no effect on the lower atmosphere.

We note that while the deposition of heat in the lower atmosphere pushes the radiative-convective boundary to larger pressures, thus preventing the exoplanet from losing its primordial entropy quickly \citep{gs02}, this effect can only explain observed radii up to about 1.3 Jupiter radii.  To explain the existence of hot Jupiters with larger radii, the dissipation of heat in the deep, convective interior is required \citep{burrows07,baty11}.  Thus, even temperature-pressure profiles which dissipate most of the heat superficially ($\gamma_0>1$) will still act to keep the exoplanet inflated because the deep isotherm corresponds to a non-negligible amount of electrical conductivity.

We conclude that the variations in the absorption opacities ($\gamma_0$) and albedo (${\cal A}$), as well as the cloud/haze properties, naturally account for a range of values of the Ohmic power dissipated at depth ($\sim 10$ bar).  If Ohmic dissipation is the dominant mechanism for inflating hot Jupiters, then this implies a scatter in the measured radii at a given strength of stellar irradiation.

\section{Discussion}
\label{sect:discussion}

\subsection{Connections to Previous Work}

We have developed models which elucidate the link between atmospheric scattering, absorption and Ohmic dissipation in hot Jupiters at a given strength of stellar irradiation.  Our study is complimentary to that of \cite{menou12}, who developed analytical scaling laws for the variation of magnetic drag and Ohmic dissipation with the strength of stellar irradiation, while we have explored the effects of stellar irradiation on heat redistribution and energy dissipation (including Ohmic dissipation), via numerical simulations, in a companion paper \citep{php12}.  Our simplifying assumption for the atmospheric dynamics is corroborated by the results of these simulations.  Furthermore, our work is also complimentary to that of \cite{wu12}, who adopted a two-zone model consisting of an isothermal envelope (i.e., a constant temperature-pressure profile) surrounding an isentropic core.  The variety of temperature-pressure profiles used in the present study highlights the importance of atmospheric effects on the evolution of a hot Jupiter, as the atmosphere provides a boundary condition for the evolution of the interior (e.g., \citealt{burrows03,burrows11}).

\subsection{Evolutionary Effects}
\label{subsect:evo}

We have not examined effects stemming from the evolution of a hot Jupiter, which we will now discuss.  Firstly, an increased atmospheric opacity slows down the gravitational contraction and internal cooling of the exoplanet, as the flux emanated scales inversely with the opacity \citep{ab06,burrows07}.  Such a retardation or suspension of the cooling has been studied by \cite{wu12}, who showed that Ohmic dissipation cannot effectively re-inflate a hot Jupiter after it has cooled.  Secondly, an enhanced metallicity --- whether in the form of a core or distributed throughout the envelope --- implies an enhanced surface gravity and a smaller radius of the exoplanet \citep{mf11}.  In turn, these effects imply that the amount of dissipation required to explain an observed radius may be orders of magnitude larger compared to a situation with lower metallicity \citep{boden01,boden03,burrows07}.  In other words, a hot Jupiter possessing a core or a metal-enriched atmosphere tends to be less inflated, all else being equal.  Thirdly, the presence of a temperature inversion itself will lead to smaller radii purely as an evolutionary effect \citep{pg11}.  All of these evolutionary effects will contribute to the scatter in the measured radii of hot Jupiters at a given strength of stellar irradiation.

\subsection{Other Observational Consequences}

The variations in atmospheric opacity are manifested in other observable ways.  At a basic level, hot Jovian atmospheres are characterized by two time scales, namely the radiative ($t_{\rm rad}$) and advective ($t_{\rm adv} \sim R/c_s$) time scales \citep{sg02,ca11}.  It is a generic statement that the radiative time scale \citep{gy89},
\begin{equation}
t_{\rm rad} \sim \frac{c_P P}{\sigma_{\rm SB} g T^3},
\end{equation}
increases with depth or pressure, such that the atmosphere is dominated by advection at large pressures.  (The specific heat capacity at constant pressure is denoted by $c_P$; its value is 14550.4 J K$^{-1}$ kg$^{-1}$ for $\Gamma=7/5$ and $m=2m_{\rm H}$.)  At the top of the atmosphere ($\lesssim 1$ mbar), we have $t_{\rm rad} \ll t_{\rm adv}$ such that the atmosphere is predominantly radiative.  If starlight is mostly deposited at the top of the atmosphere, then heat redistribution from the day- to the night-side hemisphere is inefficient; if it is deposited mostly at depth, then heat redistribution becomes efficient.  

\begin{figure}
\centering
\includegraphics[width=\columnwidth]{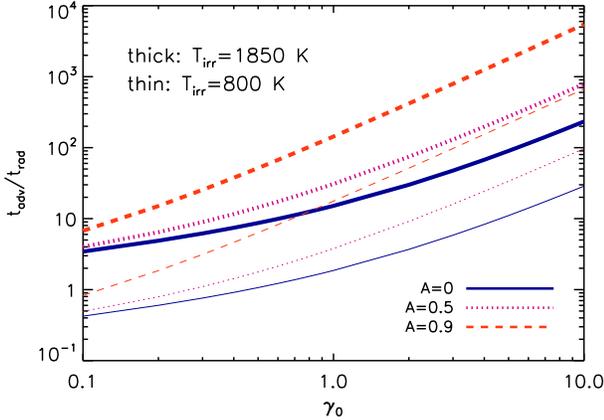}
\caption{Ratio of advective to radiative time scales, evaluated at the photon deposition depth, as a function of the optical to infrared opacity ratio $\gamma_0$ and for three values of the Bond albedo ${\cal A}=0, 0.5$ and 0.9.  The broadband, infrared absorption opacity is fixed at $\kappa_{\rm L} = 0.01$ cm$^2$ g$^{-1}$.  The atmosphere becomes more radiative (or less advective) as $t_{\rm adv}/t_{\rm rad}$ increases.}
\label{fig:ratio}
\end{figure}

In the presence of scattering, starlight is predominantly absorbed at the photon deposition depth \citep{hhps12},\footnote{In the presence of scattering, the photon deposition depth always sits deeper than the optical photosphere.  In the absence of scattering, the two layers are coincident.}
\begin{equation}
P_{\rm D} \approx \frac{0.63 g}{\kappa_{\rm S}} \left( \frac{1-{\cal A}}{1+{\cal A}} \right).
\end{equation}
It is thus plausible to evaluate $t_{\rm adv}/t_{\rm rad}$ at $P=P_{\rm D}$.  Figure \ref{fig:ratio} shows that $t_{\rm adv}/t_{\rm rad}$ is in general an increasing function of $\gamma_0$.  For a fixed broadband, infrared absorption opacity $\kappa_{\rm L}$, an increasing value of $\gamma_0$ corresponds to an increasing value of the broadband, optical absorption opacity $\kappa_{\rm S}$ and thus a decreasing value of $P_{\rm D}$.  At a given value of $\gamma_0$, increasing the Bond albedo ${\cal A}$ also decreases $P_{\rm D}$, since the overall penetration depth of an optical stellar photon, in the vertical direction, decreases when scattering is present.

As the photon deposition depth shifts to higher altitudes, it samples an increasingly radiative part of the atmosphere --- advection becomes more sluggish and heat redistribution becomes less efficient.  The angular shift of the peak, from the substellar point, of the photospheric infrared flux emanating from the hot Jupiter --- known as the ``hotspot offset" --- is a proxy for the efficiency of heat redistribution, with smaller shifts corresponding to less efficient redistribution.  It is thus natural to expect that --- all other things being equal (e.g., $g$, $\kappa_{\rm L}$) --- hot Jovian atmospheres possessing temperature inversions, at a given strength of stellar irradiation, will have less efficient heat redistribution and a higher day- to night-side flux contrast, an expectation which has been corroborated by the atmospheric circulation simulations of \cite{php12}.  While these simulations do not include scattering in the optical, we expect atmospheres with non-zero albedos to produce a similar effect.  Ultimately, stellar irradiation remains the main driver of heat redistribution and energy dissipation with opacity effects playing a secondary role \citep{php12}.

\section{Conclusion}
\label{sect:conclusion}

Collectively, the opacity effects (absorption and scattering) we have discussed, which may be attributed to variations in the metallicity and/or cloud/haze properties, as well as expected variations in the magnetic field strength, naturally introduce a scatter to the measured radii of hot Jupiters at a given strength of stellar irradiation \emph{if} Ohmic dissipation is the dominant mechanism for maintaining radius inflation.  Furthermore, the evolutionary effects we discussed in \S\ref{subsect:evo} will also contribute a scatter to the measured radii and it remains to be determined which set of effects provides the dominant source of scatter.  Future multi-wavelength measurements of thermal phase curves, from a larger sample of hot Jupiters, will disentangle these effects.

\vspace{0.1in}
\textit{KH acknowledges generous support by the Zwicky Prize Fellowship of ETH Z\"{u}rich (Star and Planet Formation Group; PI: Michael Meyer).  He is grateful to Christiane Helling, Fr\'{e}d\'{e}ric Pont, Nick Cowan, Adam Burrows and an anonymous referee for constructive comments which improved the clarity and quality of the manuscript.}


\label{lastpage}

\end{document}